# Direct observation of photonic spin Hall effect in Mie scattering

Aizaz Khan[1,2,†], Nikolay Solodovchenko[3,†], Dongliang Gao[1*], Denis Kislov[4], Xiaoying Gu[1], Yuchen Sun[5], Lei Gao[1,2*], Cheng-Wei Qiu[6*], Alexey Arsenin[4], Alexey Bolshakov[4], Vjaceslavs Bobrovs[7], Olga Koval[8], Alexander S. Shalin[2,4,9,10*]

[1] School of Physical Science and Technology, Soochow University, Suzhou 215006, China & Jiangsu Key Laboratory of Frontier Material Physics and Devices, Soochow University, Suzhou 215006, China.

[2] School of Optical and Electronic Information, Suzhou City University & Suzhou Key Laboratory of Biophotonics & Jiangsu Key Laboratory of Biophotonics, Suzhou 215104, China.

[3] School of Physics and Engineering, ITMO University, Saint-Petersburg 197101, Russia.

[4] Moscow Center for Advanced Studies, Kulakova str. 20, Moscow 123592, Russia.

[5] School of Electrical and Electronic Engineering, 50 Nanyang Avenue, Nanyang Technological University, Singapore, 639798, Singapore.

[6] Department of Electrical and Computer Engineering, National University of Singapore, Singapore 117583, Singapore.

[7] Riga Technical University, Institute of Photonics, Electronics and Telecommunications, 1048 Riga, Latvia

[8] National Research Center "Kurchatov Institute", 1 Kurchatov Square, 123182, Moscow, Russia

[9] Centre for Photonic Science and Engineering, Skolkovo Institute of Science and Technology, Moscow, 121125, Russia.

[10] Beijing Institute of Technology Beijing 100081, China

Emails: DL Gao: dlgao@suda.edu.cn; L Gao: leigao@suda.edu.cn; C-W Qiu: chengwei.qiu@nus.edu.sg;

A Shalin: alexandesh@gmail.com

**Abstract:**

The photonic spin Hall effect (PSHE), a hallmark of spin–orbit interaction of light, has long been considered a promising route toward spin-controlled functionalities in nanophotonics. Yet, its practical realization has been severely limited by the inherently weak spin–orbit coupling in typical systems, resulting in vanishingly small transverse shifts and extremely low scattering efficiency. This fundamental trade-off has rendered the PSHE observable only through complex weak measurement protocols and signal amplification — approaches that come at the cost of further intensity loss, particularly in nanoscale systems. In this work, we overcome this longstanding challenge by introducing a novel mechanism based on symmetry breaking and mode coupling in a standalone scatterer, which unlocks a regime of Friedrich-Wintgen superscattering with strong near-field spin–orbit interaction. This allows for simultaneous enhancement of both the photonic spin Hall shift and the far-field scattering intensity — boosting the latter by nearly two orders of magnitude compared to conventional dipolar particles. Through tailored multipolar interference, the PSHE is made accessible at experimentally convenient angles, enabling post selection-free detection. We report the first direct experimental observation of the PSHE from a single superscattering particle, achieved in the microwave regime via polarization-resolved far-field measurements. Our findings not only validate a new physical pathway for enhancing spin-dependent light–matter interactions, but also establish a robust, scalable platform for spin-based photonic technologies. This breakthrough opens new avenues in precision optical metrology, advanced imaging, LIDAR systems, and integrated photonic circuitry, bridging a critical gap between fundamental spin optics and real-world applications.

**Introduction**: The spin Hall effect is probably one of the most striking effects in electronic or photonic systems. The effect originates from the spin-orbit interaction (SOI) of electrons or photons[1, 2]. Akin to the spin Hall effect in electronic systems[3], the photonic spin Hall shift (PSHS) refers to a transverse beam shift of the reflected, refracted or scattered light[4, 5]. While the shift of photons has been predicted and explained in theory for more than 90 years[6], observation of PSHS in scattering from a standalone particle is still challenging because the SOI of light is usually very weak. For the PSHS at a planar interface, quantum weak measurements[7, 8] or multiple reflections[9, 10] are exploited to amplify and detect tiny shifts in experiments, which arises intrinsically by plummeting the intensity of incident polarized state. Hence, to observe the spin-dependent shift, enhanced PSHS with high efficiency is most desired, which for traditional interfaces can be realized by using metamaterial and metasurface[11, 12, 13]. Recently, an all-dielectric metasurface has been proposed to simultaneously enhance the shift and efficiency of PSHS in the optical regime[14].

As the geometry scales down to the level of a single nanoparticle, in addition to weak spin-orbit coupling, the intrinsic dilemma between the PSHS and intensity becomes adverse and presents a significant obstacle to observation of PSHS in the experiment, even at normal incidence. For example, the PSHS from spherical nanoparticles can be greatly enhanced in dual symmetric systems[15, 16].

However, owing to the inherent inverse relationship between the PSHS and the scattering intensity, the enhancement of PSHS arises at wavelengths attributed to Kerker conditions[17], where destructive interference suppresses scattering intensity to nearly zero at the precise angles corresponding to maximum PSHS. As illustrated in Fig. 1c, signal amplification through post selection inherently comes at the cost of further reducing intensity[18], making it extremely challenging to apply in measuring such subtle optical localization errors in the position of a standalone scatterer and eliminates the possibility of practical applications. Thereafter, the direct observation of PSHS from a standalone particle elevates the need for simultaneous enhancement of near-field SOI and far-field scattering efficiency. The SOI of light interacting with a low-index particle can be boosted in the near-field by replacing it with a high-index one[19], such as Fig. 1b. Multipolar engineering of high index particles has also enabled the spatial and temporal control of light and has opened a new field of resonant meta-photonics known as Mie-tronics[20, 21, 22]. A theoretical approach based on morphology optimization of high-index particle supporting strong near-field SOI that leads to efficient PSHS was realized by overlapping the electric and magnetic modes at their resonance[23] rather than their tails. Despite a nearly one-order enhancement in scattering intensity, it remains too weak, and, moreover, its detection is further hindered by experimental constraints—most notably, the positioning of the excitation antenna, which obstructs the scattered light and presents a shadowing effect for the detecting waveguide at the large angles where the PSHS reaches its maximum. Therefore, the photonic spin Hall effect remains largely a fundamental phenomenon, with practical applications still out of reach.

Here, we show that Friedrich-Wintgen mechanism of superscattering in subwavelength 3D objects[24] can greatly boost the efficiency of far-field scattering when the SOI in the near-field is enhanced as well. Compared to the intuitive method that increases the scattering intensity by overlapping the electric and magnetic modes at their resonances, here, the modes interference mechanism (governing bound states in the continuum in standalone particles) pushes involved multipoles and the overall scattering cross-section (SCS) above the single channel limit[25, 26]. By careful modes engineering, power exchange becomes possible, which results in super-dipolar modes from a nanoparticle with broken spherical symmetry. Therefore, the scattering intensity is increased by almost two orders of magnitude at the angles of enhanced PSHS. Moreover, the slope of the azimuthal component of the Poynting vector turns positive, leading to pronounced helical flow and hence the scattering angles of the maximum shift (around 124°) are not close to the backward direction, as that (around 160°) in conventional dual systems[15], which facilitates the direct observation of PSHS. Our first-to-date post selection-free microwave experiment shows that the detected shifts and scattering intensity agree well with the predicted ones by theory. This newly uncovered mechanism of the photonic spin Hall shift holds great promise for a wide range of applications, paving the way for next-generation RADAR and LIDAR systems as well as advanced photonic circuitry.

**Results:**

We start our analysis by overviewing the problem of low intensity $i(\theta)$ for enhanced PSHS. When an incident light in a pure state is scattered by a particle, the helicity-dependent spiralling power flow results in a shift of the far-field scattered intensity, as illustrated in Fig. 1a. This effect is manifested as a transverse shift of the perceived location of the particle. The difference between the intensity maxima is always twice the shift in the position. This PSHS is perpendicular to the scattering plane and arises as a consequence of the conservation of angular momentum. For a finite-size particle, the magnitude of the shift can be obtained in the far-field by the pure geometric argument, beyond the dipolar regimes as[27, 28],

$$\Delta_{SH} = \lim_{r\to\infty} r(-S_\phi / S_r)\hat{\phi}, \tag{1}$$

where $r$ is the radial distance, $S_\phi$ and $S_r$ are the azimuthal and radial components of the Poynting vector $\mathbf{S}$, respectively. An intuitive approach to enhanced PSHS is to simultaneously excite the electric and magnetic dipoles with the same magnitude, which is theoretically realized in particles satisfying electromagnetic dual symmetry[15]. However, dual symmetry for spherical scatterers usually requires the first Kerker condition[17], where the scattering coefficients of electric and magnetic modes should be exactly equal and in phase[29]. On the other hand, the intensity $i(\theta)$ is proportional to the amplitude square of the scattering coefficients (see supporting information). Thus, it is necessary for the PSHS to be enhanced simultaneously with a high far-field intensity at the angle of the enhanced PSHS if they interfere at their resonances rather than the tails. Morphology optimization of dual scatterers provides one route to achieve high intensity while preserving the enhanced magnitude of PSHS[23]. However, there is a caveat rooted in the first Kerker condition: due to destructive interference, the PSHS tends to enhance in the backscattered direction with a resonance at the tail of multipoles, where the far-field intensity faces an inherent dilemma of plummeting to zero in the case of conventional targets[23]. This intrinsic trade-off renders conventional weak measurements unsuitable for observing PSHS from a standalone particle, where additional intensity loss arises while amplifying the tiny shift (as illustrated in Fig. 1c). However, the proposed approach pushes the scattering from a standalone particle to achieve both high intensity and PSHS in the far-field together with observation directions free from shadowing effects, and strong SOI in the near-field.

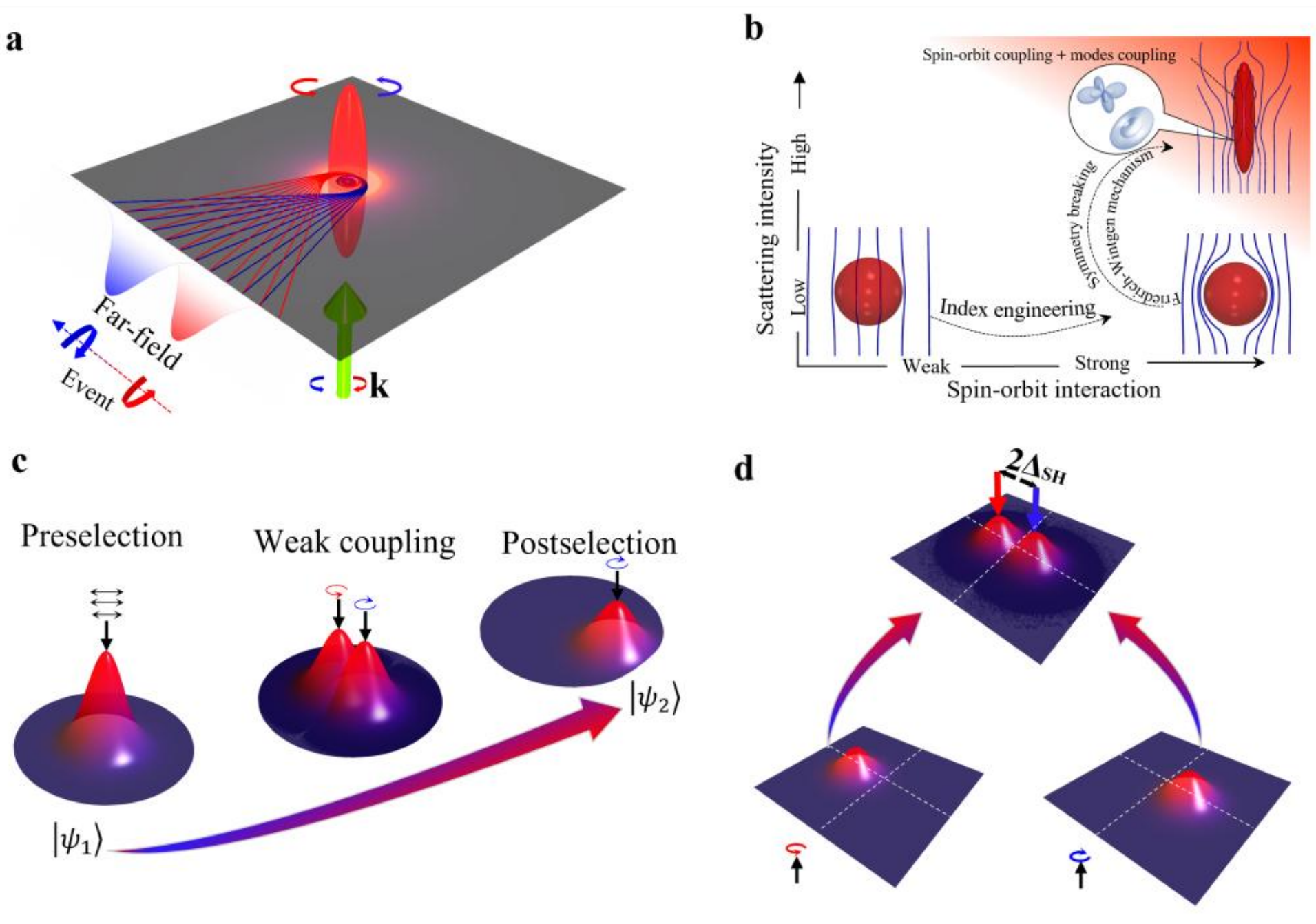

**Figure 1 | Artistic view of the post selection-free observation of the spin Hall effect by simultaneous spin-**

**orbit and modes coupling. a** Schematics of helicity-dependent vortices arising due to strong spin-orbit coupling exhibited by the scattered Poynting vector upon excitation by light in a pure state of polarization. The rotation direction of the vortex depends on the incident helicity due to which the scattered intensity at a particular observation direction in the far-field is shifted in each event by an equal amount, but in opposite directions. **b** Illustration of the trade-off between the near-field spin-orbit interaction (photonic spin-Hall shift) and the far-field scattering intensity. The weak spin-orbit interaction in low-index particles is now brought to the regime of strong coupling through index and shape engineering i.e., by replacing the particle with high-index particle with broken spherical symmetry. In addition, modes' coupling via Friedrich-Wintgen mechanism is induced together with the strong spin-orbit interaction, which results in simultaneous enhancement of the spin-Hall shift and scattering intensity. **c** Artistic view of weak measurements and intensity loss. **d** Schematic of the post selection-free superscattering induced spin Hall shift for opposite incident helicities.

To achieve these simultaneously, we break the spherical symmetry of the high-index (e.g., silicon in visible and IR spectral ranges) particle to achieve modes' coupling via Friedrich-Wintgen mechanism, as illustrated in Fig 1b. Such symmetry breaking leads to superscattering, which enhances light-matter interaction by breaking the single channel limit by forming the super multipoles[25] and, at the same time, via the spectral overlap of several multipole resonances[24, 30]. Without the loss of generality, we process with left circularly polarized light (LCP) defined as a superposition of $x$ and $y$-polarized light propagating in the $z$-direction, i.e., $\mathbf{E}^{inc} = e^{i\mathbf{k}z}(\hat{x} - i\hat{y})$, impinges on a high-index particle (to induce strong SOI) with broken rotational symmetry. To obtain the SCS, intensity as well as the PSHS, the scattered electric $\mathbf{E}^{sca}$ and magnetic $\mathbf{H}^{sca}$ fields are expanded into the series summation form of the spheroidal vector wave functions $\mathbf{M}^{(3)}_{omn}$ and $\mathbf{N}^{(3)}_{omn}$ [31].

$$\mathbf{E}^{sca} = \sum_{n=1}^{\infty} i^n \left( i\alpha_n \mathbf{N}^{(3)}_{emn} - \beta_n \mathbf{M}^{(3)}_{omn} + \alpha_n \mathbf{N}^{(3)}_{omn} - i\beta_n \mathbf{M}^{(3)}_{emn} \right), \tag{2}$$

$$\mathbf{H}^{sca} = \frac{k}{i\omega\mu} \sum_{n=1}^{\infty} i^n \left( i\alpha_n \mathbf{M}^{(3)}_{e1n} - \beta_n \mathbf{N}^{(3)}_{o1n} + \alpha_n \mathbf{M}^{(3)}_{o1n} - i\beta_n \mathbf{N}^{(3)}_{e1n} \right), \tag{3}$$

where the coefficients of the incident, scattered and internal fields can be obtained by applying the boundary conditions[31]. As the incident field propagates normally along the $z$-axis, the subscript $m$ can be taken as unity, whereas the $o$ and $e$ represent odd and even dependence corresponding to $\sin(m\phi)$ and $\cos(m\phi)$. Note that the expressions and calculations are all in the spheroidal coordinate system; one can easily transform the fields into spherical coordinates.

At this stage, by using scattered and incident fields, we obtained the total SCS, which is defined as $\sigma = \lim_{r\to\infty} 4\pi r^2 |\mathbf{E}^{sca}|^2 / |\mathbf{E}^{inc}|^2$ (see supplementary information). To achieve superscattering, the total SCS must greatly surpass the maximal SCS of a dipolar particle with spherical symmetry, the so-called single channel limit[30]. Therefore, the maximum contribution to SCS by each multipole is bound to $\sigma_{\max} = (2n+1)\lambda^2 / 2\pi$ for a standalone particle in free space, where $n$ is the scattering channel, i.e., $n=1$ for dipoles, $n=2$ for quadrupoles, and so on. Thus, arbitrarily large total SCS can be obtained by the conventional mechanism of superscattering - maximizing the contributions of sufficient number of channels (multipoles)[24] at the same frequency. On the other hand, the broken spherical symmetry can induce modes' coupling, provided that a single or several multipoles receive contributions from other multipoles and resulting in the formation of super multipoles[25]. Thus, the total contribution from a single multipole to the SCS is no longer limited to the single-channel limit as explained by the Friedrich-Wintgen mechanism[25, 32]. In the following, we show that for some particles it is possible to enhance scattering by both mechanisms.

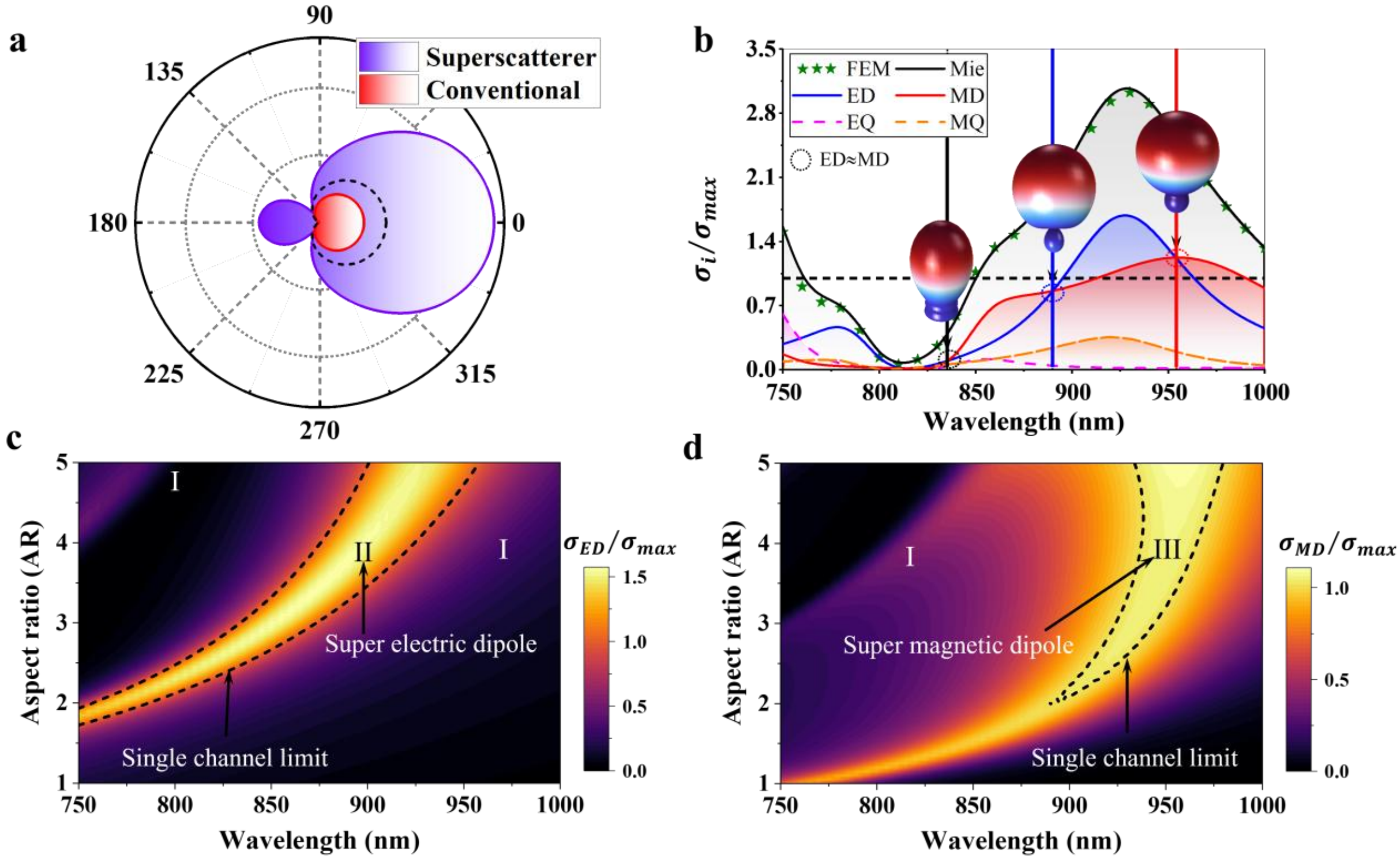

**Figure 2 | Superscattering emerging via Friedrich-Wintgen Mechanism. a** The far-field radiation patterns suitable for enhancing PSHS. The red loop represents conventional scatterers at the first Kerker condition bounded by single-channel limit (black dashed line). The blue loop represents superscatterer near the Kerker criteria with enhanced scattering far beyond the single channel limit at incident wavelength 890 nm. **b** Partial contributions of constituent's multipoles as a function of incident wavelength. Super dipoles contribute to total scattering cross-section and drive superscattering (exceeding single-channel limit by three times). The far-field radiation patterns at various wavelengths corresponding to the overlapping electric and magnetic dipoles are presented as insets. The superscatterer is defined by its aspect ratio (AR, defined as *a/b*), with *a* being the semi-major axis (500 nm) that is aligned in the incident direction of the wave, and the semi-minor axis *b* (100 nm). **c**, **d** The smooth deformation of a sphere (radius = 100 nm) to a prolate spheroid induces mode coupling and gives rise to **c** super electric (region II) and **d** super magnetic (region III) dipoles that surpass the fundamental dipolar limit (black dashed line) in a broad spectral region. In region I, both electric and magnetic dipoles are lower than the single-channel limit. The refractive index of the particle is taken as that of silicon, and the particle is embedded in air.

To establish a consistent comparison of scattering enhancement, we normalized the scattering cross-section of each multipole by the corresponding maxima defined by $\sigma_{\max} = (2n+1)\lambda^2 / 2\pi$. If $\sigma_i$ is the scattering cross-section due to a single channel, then the quantity $\sigma_i / \sigma_{\max}$ is usually less than 1 for a conventional scatterer, with a maximum possible value that can be achieved under the condition $\sigma_i = \sigma_{\max}$ is 1. In what follows, super electric and super magnetic dipoles emerge by breaking the rotational symmetry of the sphere in the direction parallel to the propagation direction to form a prolate spheroid, as illustrated in Fig. 2b. The contribution from super electric and super magnetic dipoles greatly enhances scattering and pushes the total SCS beyond the single channel limit, as shown in Fig. 2b. To confirm the results of generalized Lorentz-Mie theory, we performed simulations based on the finite element method (FEM) implemented in COMSOL Multiphysics. Both theory and simulations agree well with each other. By maintaining a constant equatorial radius and changing the polar radius of the prolate spheroid through the variation of the aspect ratio, both super electric and super magnetic dipoles emerge around AR~1.5, whereas both the super electric and super magnetic dipoles vanish as the AR approaches unity (sphere), as illustrated in Fig. 2c,d. This amplification of the total SCS is primarily due to the enhanced contribution from the super multipoles in that wavelength range. Therefore, by breaking the spherical symmetry, we have achieved superscattering in a standalone 3D scatterer via super multipoles and the accidental overlap. By being able to enhance the contribution of dipolar channels and interfering them near their resonances in the superscattering regime, one can enhance light-matter interaction to bring the intensity at the peak PSHS with strong spin-orbit coupling, to a detectable level in stark contrast with conventional dual particles[15, 16, 23], as demonstrated in Fig. 2a. In the following, we show that such superscatterer drastically boosts the intensity at the observation angle suitable for an experimental observation of the PSHS.

In the near-field, strong SOI results in the far-field PSHS, which merely depends upon the scattering angle $\theta$, provided that the scatterer is a dipole[33]. However, for a finite-sized particle, the PSHS depends on both the optical properties of the particle and the scattering angle. Therefore, the enhanced PSHS arises at a certain scattering angle $\theta$ accompanied by strong SOI, which is mediated by the large discrepancy of the optical constant at the air-particle interface, resulting in a large optical potential gradient[27]. Generally, the mutual relation between the PSHS and the far-field intensity is inverse, with a similar analogy in 2D cases. At the level of a single particle, the intensity is tied to the radial component of the Poynting vector $S_r$. Meanwhile, the PSHS is enhanced if the radial component $S_r$ is small as a direct consequence of Eq. 1. A parallel comparison can be made in planar interfaces, where the reflectivity of incident polarization is greatly suppressed while enhancing that of other polarization to achieve enhanced PSHS[34]. Thus, direct observation of PSHS from a standalone particle requires a relative enhancement of the azimuthal component to ensure high efficiency.

In Fig. 3, we state the challenge in experimental detection of the enhanced PSHS at the backscattered direction for conventional cases. First, the scattered intensity is detected in a plane that is tangent to an imaginary sphere with radius *r*. At a fixed observation direction, the coordinates of this plane are defined by the dimensionless parameters $\mu = kx$ (transverse) and $\eta = ky$ (longitudinal). The continuous variation in the observation direction will lead to instantaneous tangent planes that eventually lie on the surface of the imaginary sphere enclosing the scatterer at its centre, as shown in Fig. 3a. For a conventional quasi-dual particle with spherical symmetry, the PSHS is enhanced by enforcing high directional scattering via satisfying the first Kerker conditions[15, 16]. Thus, in the backscattered direction, the destructive interference between electric and magnetic dipoles results in enhanced PSHS at the scattered angle around 160°. However, due to the overlapping dipoles at their tails, the far-field intensity is vanishingly small (red panel in Fig. 3b) and is not enough to directly detect the PSHS. Geometry optimization of conventional scatterers enforces the overlap of scattering coefficients at their resonances, leading to enhanced intensity by one order (blue-panel)[23]. Yet, with low azimuthal and radial components of the Poynting vector, the intensity is still too small, and the resonance of PSHS lies at the extreme backscattered direction which is not suitable for direct experimental observation, as marked by conventional II in Fig. 3b.

To realize simultaneous enhancement of PSHS and its far-field efficiency, the superscatterer at the incident wavelength 890 nm (solid blue line in Fig. 2b) is exploited. When the particle is in the superscattering state, the enhancement of the PSHS is observed at the angle $\sim 124°$, as shown by the black line in Fig. 3b. At the same time, the far-field intensity at the maximal PSHS is enhanced by 100 times that of the conventional spherical particle, as shown in Fig. 3b. In addition to the large radial component of the Poynting vector, the morphology of the superscatterer induces a helical flow around the particle, leading to a large azimuthal component as well in stark contrast to conventional scatterers[15] (see supporting information). This empowers both PSHS and the associated intensity at the scattering angle, and makes it suitable for direct observation. It is to be noted that, unlike earlier works of PSHS, we have assumed losses in the particle, as all the results in our work consider a realistic particle rather than only fixed permittivity with no loss. In stark contrast to conventional particles, both the enhanced far-field PSHS and the intensity are obtained for the superscatterer, and the angle of maximal PSHS is much smaller, opening a way for a plenty of applications. With preserved enhanced PSHS, the intensity is almost 100 times that of a conventional dipolar particle. These are the main hallmarks of the current work, which allow for experimental direct observation of an enhanced far-field PSHS.

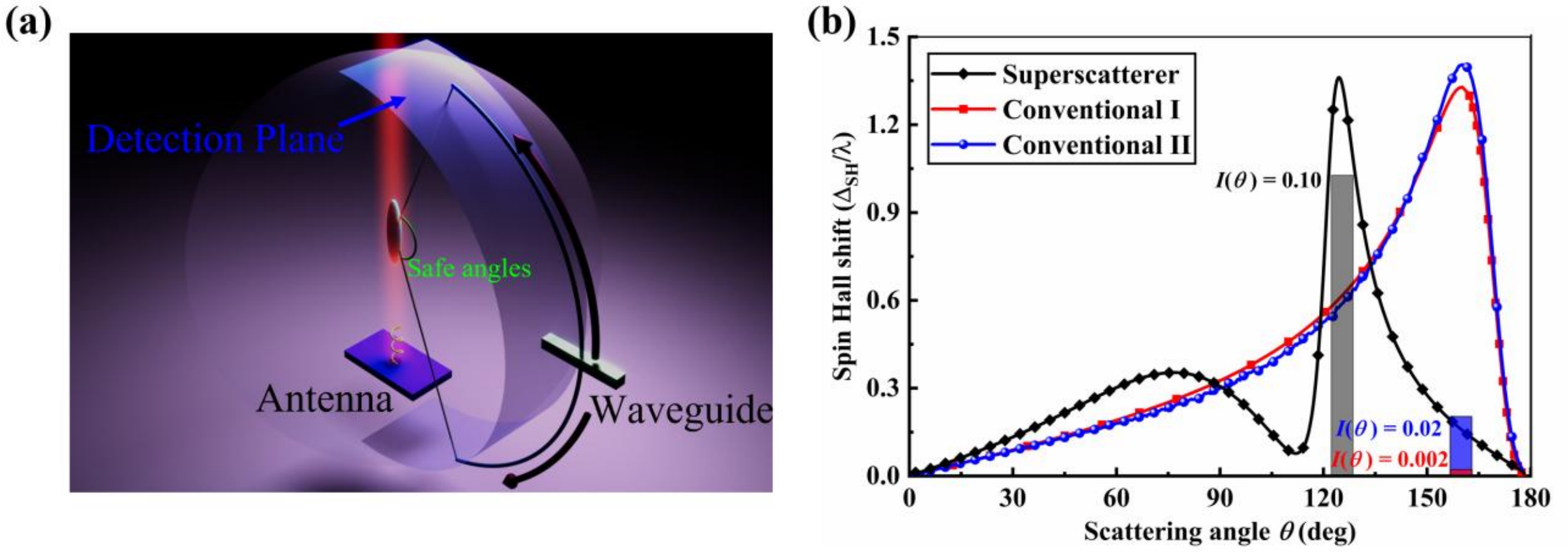

**Figure 3 | Superscattering driven simultaneous enhancement of photonic spin Hall shift and scattering intensity. a** Artistic representation of the experimental setup representing the limitations of large-angle detection. In the detection plane, the scattered field at the maximal PSHS is blocked from the detector (waveguide) by the antenna, which poses a challenge in an experimental measurement of the photonic spin Hall shift (PSHS) in conventional cases. **b** The far-field photonic spin Hall shift and scattering intensity for the particle in the superscattering regime, and conventional scatterers. The intensity for the superscatterer (gray rectangle) drastically surpasses those for the conventional scatterers (red and blue rectangles) at the maximal PSHS. For numerical calculations, the refractive index for all three cases is taken as that of silicon, and the incident wavelength is chosen 890 nm. The optimal diameter for the conventional case I is 219 nm, and semi-

major axis $a = 169$ nm with the aspect ratio of $0.5$ for the conventional case II. The superscatterer is defined by the same parameters as those of Fig. 2b. The peak PSHS lies at backscattered direction for conventional scatterers, while for the superscatter, the maximum PSHS appears at the angle $\theta \approx 124$. The incident polarization is taken as LCP.

The enhancement of PSHS in the far-field is fundamentally governed by the strength of SOI in the near-field[35, 36]. Together with a strong near-field SOI, the interference between the electric and magnetic multipoles affects the magnitude of the PSHS[15]. In the near-field, the trajectory of photons is governed by the orbital momentum density (OMD), manifesting as optical currents. Thus, we represent it as streamlines and characterize the polarization behaviour through the circular polarization degree (CPD) defined as $\mathrm{CPD} = |\varepsilon_0 \mathbf{E}^* \times \mathbf{E} + \mu_0 \mathbf{H}^* \times \mathbf{H}| / (\varepsilon_0 |\mathbf{E}|^2 + \mu_0 |\mathbf{H}|^2)$ [19]. A field in pure state (circular-polarization) is characterized by $\mathrm{CPD} = 1$, on the other hand, $\mathrm{CPD} = 0$ is associated with linear polarization. In Fig. 4, we show the OMD streamlines and CPD for the cases, where the electric and magnetic dipoles overlap (vertical solid lines in Fig. 2b). To get a better intuition on how the trajectory of light evolves in the near-field, the streamlines for OMD are traced in both two dimensions (*xy*-plane passing through the symmetry axis) and 3D, whereas the CPD is shown in two orthogonal half-spaces —one with surface normal along the *y*-axis (upper row) and the other along the *z*-axis (lower row). The stark contrast of optical constants at the surface of the high-index particle leads to rapidly changing optical potential and mediates strong SOI[19]. Due to this, the trajectory of photons in the near-field, featuring singularities, is strongly twisted, which is attributed to the low CPD in the near-field. Intriguingly, in Fig. 4b, the streamlines are significantly twisted as they reach the maximum geometric cross-section of the particle, while converging to the vortex singularity at the front of the particle. The saddle singularity at the rear of the particle (in the *xz*-plane) directs the photons in the forward, whereas these are collected by the vortex singularity at the front of the particle (as visualized in 3D). Then the photons are steered by the saddle singularity that emerges adjacent to the vortex in the forward direction (see Fig. 4b). This spin flow arises at the incident wavelength of 890 nm (blue solid line in Fig. 2b), where the super electric dipole and magnetic dipole interfere constructively in the forward direction. This incident wavelength corresponds to the enhanced PSHS in the superscattering regime. Contrary to this, overlapping electric and magnetic dipoles in the superscattering region (red solid line in Fig. 2b) and the wavelength of non-superscattering region (black solid line in Fig. 2b) exhibit weak twisted streamlines and correspondingly low drop in CPD, as shown in Fig. 4a, c.

It can be noted that the intensity will be far higher at the wavelength, where the total SCS reaches its maximum and also at the overlapping dipoles (red solid line in Fig. 2b). However, the corresponding PSHS at this wavelength will be very small primarily due to two reasons. As demonstrated in Fig. 4c, the trajectory of photons is weakly twisted with a low fall in CPD, leading to comparatively low SOI in contrast to that of Fig. 4b. Secondly, the enhancement of PSHS is tied with overlapping multipoles in phase. While the super electric and magnetic dipoles do not overlap at the wavelength where total SCS reaches its peak, they are also not in phase at the incident wavelength 950 nm. Thus, the suitable wavelength for both enhancement of PSHS and the scattering intensity corresponds to the conditions, where the multipoles overlap in phase along with strong SOI, as in Fig. 4b and the wavelength 890 nm.

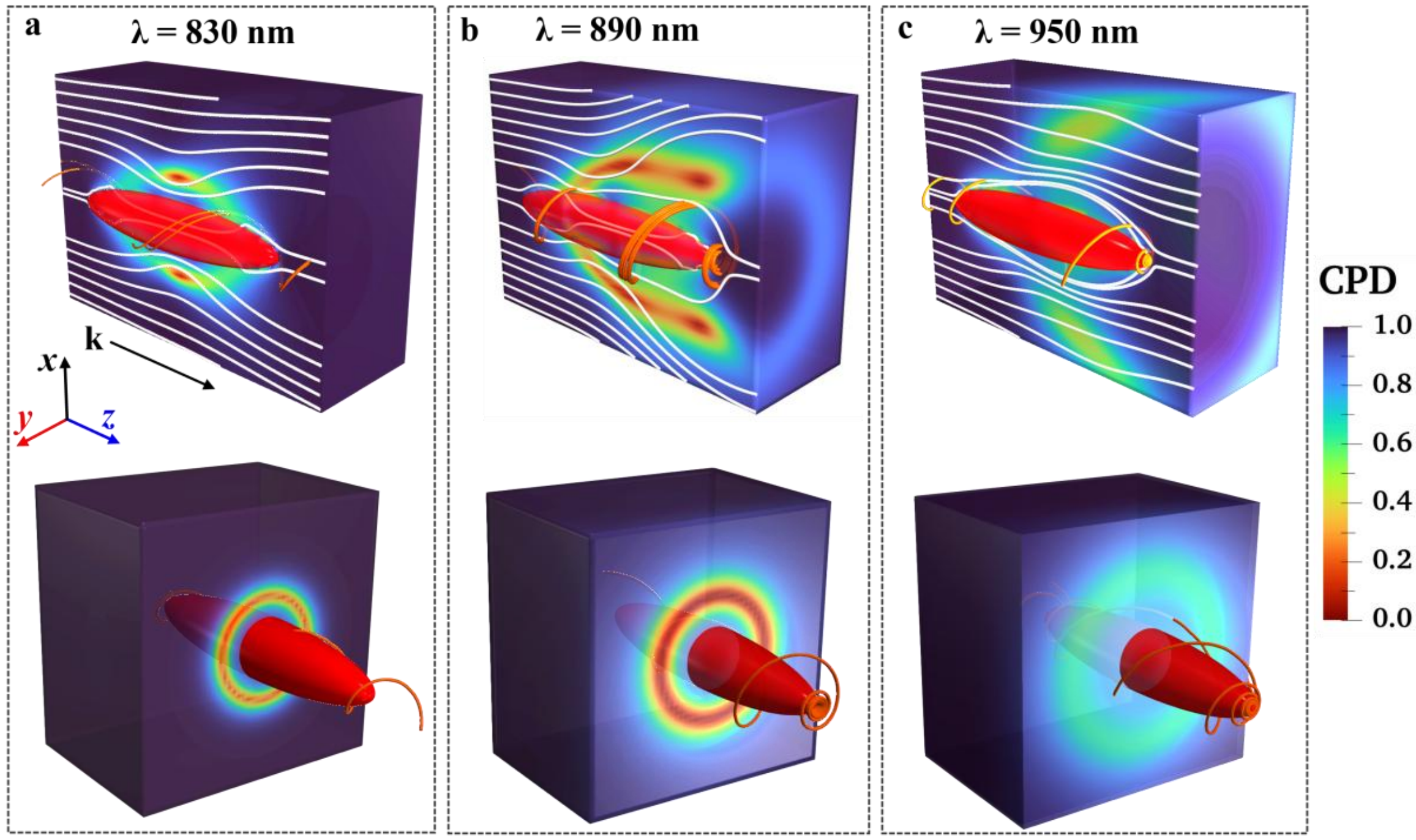

**Figure 4 | Near-field spin-orbit interaction in the superscattering particle.** The circular polarization degree of field (volume around the particle) and orbital momentum density (as streamlines or optical currents) for **a** non-superscattering regime at wavelength 830 nm and **b**, **c** superscattering regimes at wavelengths 890

nm and 950 nm, respectively, as marked in Fig. 2b. The dark red regions in the volume surrounding the particle, where the circular polarization degree drops to zero, indicate spin-free, linearly polarized field in all panels. Superscattering from high-index particle induces abrupt changes in the near-field photon trajectories, resulting in geometric phase accumulation and strong spin-orbit interaction, which is attributed to loss in the circular polarization degree. The streamlines for orbital momentum density in *xz*-plane in the upper-row are shown by the white lines, whereas they are shown in 3D by the orange colour in all panels. The parameters remained the same as in Fig. 2b except that the polarization of the incident light is LCP.

**Experimental demonstration:**

To verify our results, we performed a microwave experiment for a ceramic-based superscattering spheroid defined by semi-major axis $a = 25.995$ mm and semi-minor axis $b = 4.985$ mm, and permittivity $\varepsilon' = 15.2$ and losses $\tan(\delta) \approx 10^{-4}$. Irrespective of the geometry of the underlying structure, the shift was usually determined experimentally by Stokes' parameters[14, 37] or weak measurements[7]. Our experimental setup completely differs from the previous methods and closely resembles that of the proposed experimental setup in references[23, 38]. In our setup, circularly polarized light is incident along the spheroidal axis. The exciting antenna is located at a distance of 50 cm, and is tuned to the frequency 6 GHz. The scattered field is detected in the far-field by a waveguide with a dipole inside, the operating frequency of which is in the range of 4.9 to 7 GHz. The distance from the spheroid to the dipole is about 120 cm. To register the scattered field only from the spheroid, it is necessary to detect the scattering matrix element $S_{12}$ with and without the particle so that the difference in the measurements accounts for the removal of the background field and noise in the system. Since the rectangular waveguide detects only one polarization of the electric field along its short side, the total scattered field was obtained by combining two mutually perpendicular components. To further reduce noise, the spectra for each component were averaged over 20 measurements. Since the particle rotates along with the emitting antenna while the receiving waveguide only moves up and down along the *z*-axis, the detector is almost always oriented towards the particle. Due to complex scattering, the detection of the original position of the particle in the scattering plane is challenging, and so is the corresponding shift in the far-field. However, PSHS for RCP and LCP light have the same magnitude and opposite sign. It is then reasonable that the shift in the perceived location can be found from the difference of the maximum far-field intensity corresponding to different polarization at a particular scattering angle.

In what follows, we measure the far-field intensity for the excitation corresponding to light in the pure state of polarization (RCP and LCP) for various scattering angles in the vicinity of the peak of PSHS at the incident frequency 6.02 GHz. We determined the PSHS from the difference in the peak far-field intensity of the incident RCP and LCP waves. The results of the experiment are in a good agreement with theoretical and numerical, as shown in Fig. 5a. It should be noted that the difference in the peak intensities for RCP and LCP is always 2 times the shift in the perceived location. The experimental results must be reduced by half to get the PSHS. The far-field intensity is averaged over 20 measurements, and the result is plotted along the scattering angle, which is also in a good agreement with the theoretical prediction Fig. 5b. With $\varepsilon = \text{const}$, the experimentally obtained results can be generalized to the optical frequency range.

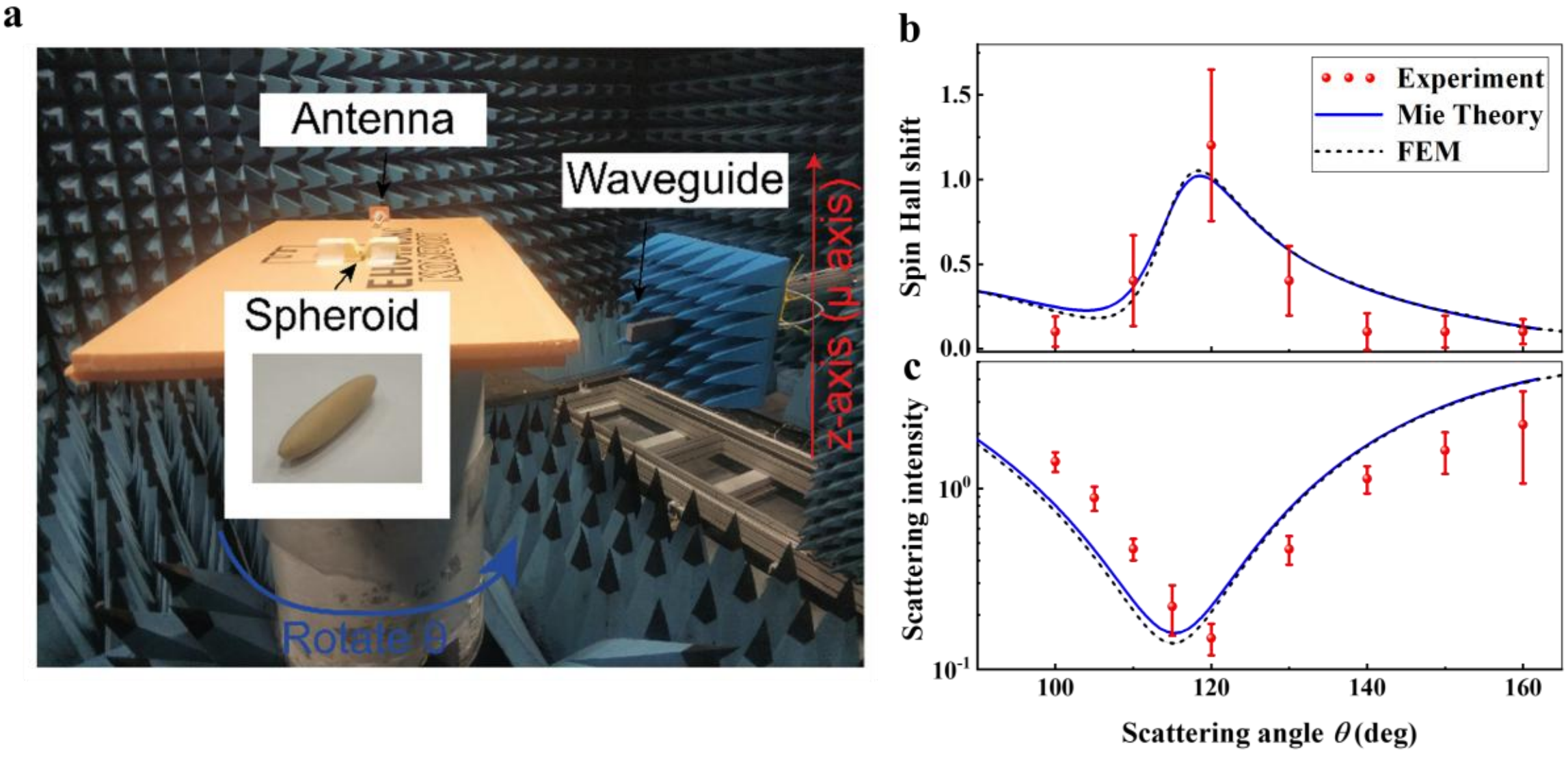

**Figure 5 | Direct experimental observation of the PSHS for standalone particle. a** Far-field view of experimental setup, showing antenna, rectangular waveguide and the ceramic particle. The exciting frequency is $6.02$ GHz, and the scatterer is made up of ceramic composite with semi-major axis $a = 25.995$ mm and semi-minor axis $b = 4.985$ mm. The permittivity of the scatterer for the theoretical curve is taken as $\varepsilon'=15.2$, losses $\tan(\delta)\approx 10^{-4}$. Comparison of the experimental results for **b** PSHS and **c** far-field intensity with the theoretical

results obtained from Generalized Lorentz-Mie theory and finite element method (FEM) based simulations. All three curves agree with each other.

**Discussions:**

We present both a theoretical prediction and a first-ever direct experimental observation of the photonic spin Hall shift (PSHS) arising from a standalone three-dimensional superscattering particle featuring strong near-field spin–orbit interaction (SOI). This result marks a significant departure from earlier studies involving dual scatterers[15], where enhancement of the PSHS came at the cost of dramatically suppressed far-field intensity — a consequence of the first Kerker condition, which renders the scattering nearly null at the very angles where the effect is strongest.

Attempts to overcome this limitation through careful morphological optimization of dual scatterers — typically by aligning multipolar resonances — have yielded only modest improvements[23]. The resulting far-field signal has remained exceedingly weak and the PSHS enhancement confined to extreme backscattering angles, making direct detection virtually impossible. In contrast, our approach leverages a fundamentally new superscattering mechanism, enabling simultaneous enhancement of both the PSHS and the far-field scattering intensity in a realistic, lossy particle[25]. Strikingly, the observed scattering intensity surpasses that of previously optimized dual scatterers with comparable PSHS by more than an order of magnitude and conventional scatterers by two orders — a key factor that makes the effect directly measurable. Moreover, the optimal scattering angle is conveniently located away from the strict backward direction, greatly facilitating experimental realization. This has allowed us, for the first time, to directly measure the PSHS in a microwave-range experiment using a single superscattering particle — a platform that is readily scalable to the optical domain. Notably, our method requires no post-selection procedures[7], further enhancing its practicality and robustness. Beyond its fundamental importance, this breakthrough opens up a wealth of exciting possibilities for real-world applications. The demonstrated platform lays a solid foundation for next-generation photonic technologies, including spin-controlled light routing, advanced RADAR and LIDAR systems, high-performance nanoantennas, and integrated spin-dependent photonic circuitry.

## Methods:

### Generalized Lorentz-Mie theory:

Scattering of electromagnetic waves from a spheroidal particle can be studied using the generalized Lorentz-Mie theory. Here we briefly introduce the main equations, the details of which can be found in the references[31, 39]. For a homogenous particle defined by the spheroidal coordinates $(\eta,\xi,\varphi)$ in an isotropic and homogenous medium illuminated by an electromagnetic wave, the time-independent parts of the fields inside and outside the spheroid particle satisfy the vector wave equations[31].

$$\nabla^2\mathbf{E}+k^2\mathbf{E}=0,\nabla^2\mathbf{H}+k^2\mathbf{H}=0, \tag{4}$$

where the propagation constant is defined as $k=k_0 n$, with $n$ being the complex refractive index at the incident wavelength $\lambda$. The solution of the above vector wave equation can be built from the separable scalar wave equation $\nabla^2\psi+k^2\psi=0$ in spheroidal coordinate systems. Considering the spheroid is defined by the coordinates of a prolate, with the use of Laplacian $\nabla^2$ in curvilinear coordinates, the resulting scalar wave equation can be written as[39].

$$\left[\frac{\partial}{\partial\eta}(1-\eta^2)\frac{\partial}{\partial\eta}+\frac{\partial}{\partial\xi}(\xi^2-1)\frac{\partial}{\partial\xi}+\frac{\xi^2-\eta^2}{(\xi^2-1)(1-\eta^2)}\frac{\partial}{\partial\phi^2}\right]\psi+c^2(\xi^2-\eta^2)\psi=0\cdot \tag{5}$$

and implementing the separable of variables techniques, i.e., $\psi_{mn}(c;\eta)=S_{mn}(c;\eta)R_{mn}(c;\xi)\Phi(\phi)$, the above equation can be split into three separate ordinary differential equations.

$$\frac{d}{d\eta}\left[(1-\eta^2)\frac{d}{d\eta}S_{mn}(c;\eta)\right]+\left[\lambda_{mn}-c^2\eta^2-\frac{m^2}{1-\eta^2}\right]S_{mn}(c;\eta)=0, \tag{6}$$

$$\frac{d}{d\xi}\left[(1-\xi^2)\frac{d}{d\xi}R_{mn}(c;\xi)\right]-\left[\lambda_{mn}-c^2\eta^2+\frac{m^2}{\xi^2-1}\right]R_{mn}(c;\xi)=0, \tag{7}$$

where the solutions of $\Phi(\phi)$ are the sin and cosine functions. The angular $S_{mn}(c;\eta)$ and radial $R_{mn}(c;\xi)$ functions, respectively satisfy the above equations. The solution to this equation can be reconstructed from the three separate second-order differential equations in $\eta,\xi$ and $\phi$, respectively, as follows.

$$\psi^{(j)}_{\substack{e\\o}mn}(c;\eta,\xi,\phi)=S_{mn}(c;\eta)R^{(j)}_{mn}(c;\xi)^{\cos}_{\sin}m\phi\,, \tag{8}$$

Where $j$ represents the kind of the radial function. For an arbitrary unit vector $\mathbf{a}$, the solution for the vector wave equation can be constructed from $\psi^{(j)}_{\substack{e\\o}mn}(c;\eta,\xi,\phi)$ ,i.e., the spheroidal vector wavefunctions.

$$\begin{aligned}&\mathbf{M}^{r}_{\substack{e\\o}mn}=\nabla\times(\mathbf{a}.\psi^{(j)}_{\substack{e\\o}mn}(c;\eta,\xi,\phi)),\\&\mathbf{N}^{r}_{\substack{e\\o}mn}=(1/k)\nabla\times\mathbf{M}^{r}_{\substack{e\\o}mn},\end{aligned} \tag{9}$$

in terms of which the fields inside and outside the particle can be expanded by using the appropriate kind of radial function. By using the boundary conditions[31], the unknown coefficients in Eq. (2) and Eq. (3) can be obtained. All the corresponding quantities can be obtained once the electric and magnetic fields are found.

**Numerical Simulations.**

To validate the results of the theory and experiment, scattering simulations under electromagnetic wave frequency domain using the finite element method (FEM) in commercial simulation software COMSOL Multiphysics© were performed. The refractive index for the particle was taken as that of silicon at the optical wavelengths to verify the results of theory. For verification of experimental results, the refractive index was calculated from the permittivity of ceramic $\varepsilon'=15.2$ while considering the tangent loss $\tan(\delta)\approx10^{-4}$ . The dimensions of the particle were taken to be the same as mentioned in the main text. Contribution to scattering from a single channel and the total scattering cross-section was calculated by the multipole decomposition method[40]. Simulations for the PSHS and scattering intensity as a function of scattering angle were calculated by using the scattered Poynting vector and the scattered fields. The air domain surrounding the particle for the calculations of PSHS and the scattering intensity was taken 4 times the incident wavelength.

**Sample fabrication.** The sample was made of ceramics, the basis of the chemical composition of which is Mg-Ti-O. Measurement showed that the permittivity of ceramics at a frequency of 6 GHz is $\varepsilon'=15.2$, $\tan(\delta)\approx10^{-4}$. Such ceramics are well amenable to mechanical action, so it can be given any complex shape, including a spheroid with specific geometric parameters. In our work, we used a rectangular ceramic parallelepiped (1 cm × 1 cm × 5.5 cm), and after mechanical processing, a superscatttering particle with semi-major axis $a=25.995$ mm and semi-minor axis $b=4.985$ mm was obtained.

**Experimental measurements.** For the experimental observation of the PSHS, we used two helix antennas, LCP and RCP. The antennas were handcrafted and tested to meet the required specifications. The antenna and the sample were fixed on a rotating table, which allows maintaining the coaxiality of the wave vector **k** of the incident field and the superscattering particle (see supplementary materials). The scattered field was detected by a rectangular waveguide, which measures only one component of the field at a time, along its short side. The antenna and waveguide were connected to a Rohde & Schwarz ZVB 20 vector analyzer. To extract the signal from a particle only, the S parameters of the system must be measured twice, with and without the particle, and then subtracted. The scattered intensity is a combination of two orthogonal components. Thus, to scan the full scattered field intensity for one polarization of the incident field, four measurements are required. To reduce noise, the measurements were averaged over 20 times.

**Data availability**

All data needed to evaluate the conclusions in this study are presented in the manuscript and the supporting information. The refractive index data can be taken as that of silicon in the desired frequency range.

## Acknowledgements

This work was supported by National Natural Science Foundation of China (12174281, 12274314, 12311530763); Natural Science Foundation of Jiangsu Province (BK20221240); Suzhou Basic Research Project (SJC2023003). The calculations of non-Hermitian superscattering are partially supported by Russian Science Foundation (23-72-00037).

## Author Contributions

These authors contributed equally: Aizaz Khan, Nikolay Solodovchenko.

## Competing interests

The authors declare no competing interests.

**Supplementary Information for**

# "Direct observation of photonic spin Hall effect in Mie scattering"

Aizaz Khan[1,2,†], Nikolay Solodovchenko[3,†], Dongliang Gao[1*], Denis Kislov[4], Xiaoying Gu[1], Yuchen Sun[5], Lei Gao[1,2*], Cheng-Wei Qiu[6*], Alexey Arsenin[4], Alexey Bolshakov[4], Vjaceslavs Bobrovs[7], Olga Koval[8], Alexander S. Shalin[2,4,9,10*]

[1] School of Physical Science and Technology, Soochow University, Suzhou 215006, China & Jiangsu Key Laboratory of Frontier Material Physics and Devices, Soochow University, Suzhou 215006, China.

[2] School of Optical and Electronic Information, Suzhou City University & Suzhou Key Laboratory of Biophotonics & Jiangsu Key Laboratory of Biophotonics, Suzhou 215104, China.

[3] School of Physics and Engineering, ITMO University, Saint-Petersburg 197101, Russia.

[4] Moscow Center for Advanced Studies, Kulakova str. 20, Moscow 123592, Russia.

[5] School of Electrical and Electronic Engineering, 50 Nanyang Avenue, Nanyang Technological University, Singapore, 639798, Singapore.

[6] Department of Electrical and Computer Engineering, National University of Singapore, Singapore 117583, Singapore.

[7] Riga Technical University, Institute of Photonics, Electronics and Telecommunications, 1048 Riga, Latvia

[8] National Research Center "Kurchatov Institute", 1 Kurchatov Square, 123182, Moscow, Russia

[9] Centre for Photonic Science and Engineering, Skolkovo Institute of Science and Technology, Moscow, 121125, Russia.

[10]Beijing Institute of Technology Beijing 100081, China

**Table of Contents:**

## S1. Relation of spin Hall shift and scattering intensity with the Poynting vector:

The photonic spin Hall shift (PSHS) in scattering is commonly associated with the two eigenmodes: left circular polarized (LCP) and right circular polarized light (RCP), which can be treated as pure states of polarization with defined helicity[1]. In a spherical coordinate system, the angular $S_\theta$ and azimuthal $S_\phi$ components of the Poynting vector are associated with transversal energy flow and are directly responsible for the apparent displacement of the position of the scatterer in both the longitudinal and the transverse directions, respectively. However, the angular component, which is independent of polarization, primarily governs the shift in the longitudinal direction and is the analog of Goos-Hänchen shift at the scale of scattering from a single particle. On the other hand, the shift in the position of the particle due to spin-orbit interaction (SOI), as perceived in the far-field, is determined by the azimuthal component, which can be calculated using Mie theory[1, 2, 3].

$$\mathbf{\Delta}_{\mathrm{SH}} = \lim_{r\to\infty} r(-S_{\phi} / S_{r})\hat{\phi}\,, \tag{S1}$$

where, $S_r$ represent the radial component of the scattered Poynting vector $\mathbf{S}$. The definition for the far-field PSHS holds true for a spherical, cylindrical or spheroidal scatterer, provided that a suitable transformation from spheroidal coordinates has been implemented. To enhance scattering via mode coupling, spherical symmetry should be broken. In such cases, the PSHS for spheroidal scatterers can be obtained by using the above definition under the far-field approximation. The corresponding far-field scattered intensity in the far-field limit can be found by using the relation.

$$i(\theta) = (kr)^2 (E_{\xi}E_{\xi}^{*} + E_{\eta}E_{\eta}^{*})\,. \tag{S2}$$

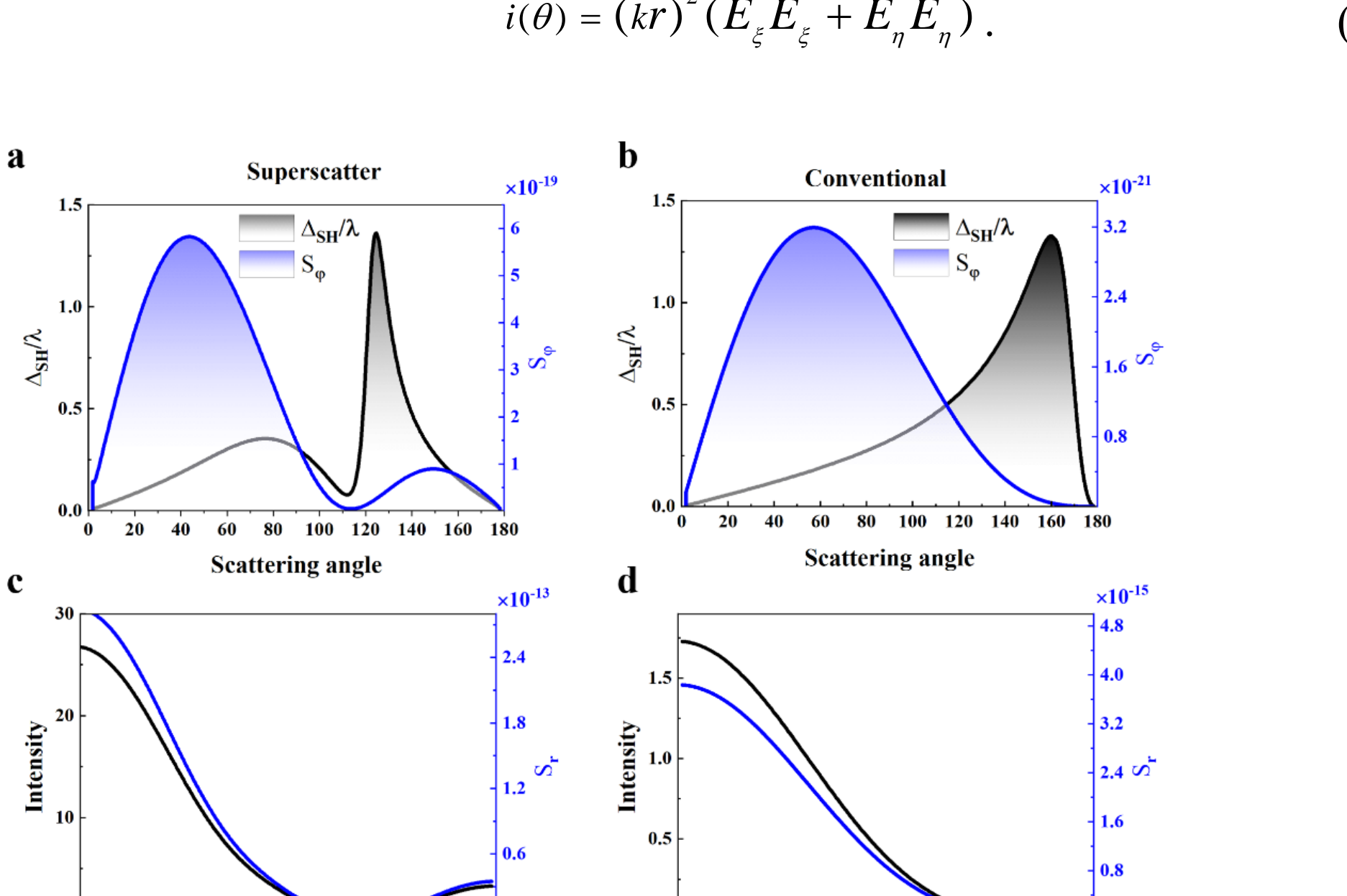

**Fig. S6| Dependence of PSHS and scattering intensity on azimuthal and radial components of Poynting vector:** PSHS and the azimuthal component of the Poynting vector for **a** superscatterer **b** conventional spherical scatterer with diameter 219 nm. The azimuthal component at the maximal PSHS is enhanced rather than sufficiently suppressed radial component for the superscatterer. The far-field scattered intensity fallows the same

trend as the radial component of the Poynting vector. Scattered intensity and the radial component of Poynting vector along the scattering angle for **c** superscatterer and **d** conventional scatterer. The superscattering particle is a prolate spheroid with semi-major axis a=500 nm and semi-minor axis b=100 nm. The incident wavelength is 890 nm.

Although, the radial component of the scattered electric field vanishes in the far-field, the scattering intensity exhibits the same variations as $S_r$, as shown in Fig. S1a. According to the definition of PSHS, enhancing PSHS in higher dimensions requires an increase in the ratio of the azimuthal to the radial component. While the azimuthal component typically has a small magnitude, a depth in the $S_r$ at a certain scattering angle can boost the PSHS, leading to lower scattering intensity at the maximal PSHS. This is what commonly observed at the extreme backscattered direction in scattering from conventional scatterers, as shown in Fig. S1. In stark contrast to this, due to enhanced scattering and helical flow, both the radial and azimuthal components of the Poynting vector are boosted by at the angle of maximal PSHS, as shown in Fig. S1. Thus, by maximizing the light-matter interaction through superscattering, one can obtain enhanced scattering intensity with preserved PSHS, provided that strong spin-orbit interaction appears in the near-field together with overlapping multipoles in phase, as we shall discuss in the following sections.

**S2. Derivation of explicit relation between PSHS and scattering intensity for spheroidal particle:**

Without the loss of generality, we assume a spheroidal superscatterer, which is governed by its semi-major axis $a$, semi-minor axis $b$ and focal length $d$ and is illuminated by LCP light propagating along the *z*-axis, i.e., $E^{inc} = e^{ikz}(\hat{x} - i\hat{y})$. For brevity, the propagating direction is taken parallel to the semi-major axis $a$ of the spheroidal superscatterer. The size parameter $q$ is defined as $q = ka$, where $k$ is the free-space wavevector. By employing the components of scattered electric and magnetic fields, one can obtain the corresponding components of the Poynting vector. The incident electric and magnetic fields can be expressed in terms of spheroidal vector wave functions $\mathbf{N}_{emn}^{(i)}$ and $\mathbf{M}_{emn}^{(i)}$ [4],

$$\mathbf{E}^{inc} = \sum_{n=1}^{\infty} i^n \left( g_n \mathbf{M}_{omn}^{(1)} - i f_n \mathbf{N}_{emn}^{(1)} + i g_n \mathbf{M}_{emn}^{(1)} - f_n \mathbf{N}_{omn}^{(1)} \right), \mathbf{H}^{inc} = \frac{k}{i\omega\mu} \sum_{n=1}^{\infty} i^n \left( g_n \mathbf{N}_{omn}^{(1)} - i f_n \mathbf{M}_{emn}^{(1)} + i g_n \mathbf{N}_{emn}^{(1)} - f_n \mathbf{M}_{omn}^{(1)} \right). \quad \text{(S3)}$$

Where the coefficients of the incident field are related, $g_n = f_n$ and can be defined as,

$$g_n = \frac{2}{\sum_{r=0,1}^{\infty}{}' \frac{2(r+2)!}{(2r+3)r!}(d_r^{mn})^2} \sum_{r=0,1}^{\infty}{}' d_r^{mn}, \tag{S4}$$

where $d_r^{mn}$ is the expansion coefficient related to the spheroidal coordinates system, whereas the prime in the summation shows that the summation is only over even values, if the difference $n-m$ is even, otherwise, it is only over odd[4]. Under normal incidence, $m$ can be taken as unity. To derive the PSHS, one needs the scattered electric and magnetic fields, which can be expressed in terms of the spheroidal vector wave functions.

$$\mathbf{E}^{sca} = \sum_{n=1}^{\infty} i^n \left( i\alpha_n \mathbf{N}_{emn}^{(3)} - \beta_n \mathbf{M}_{omn}^{(3)} + \alpha_n \mathbf{N}_{omn}^{(3)} - i\beta_n \mathbf{M}_{emn}^{(3)} \right), \mathbf{H}^{sca} = \frac{k}{i\omega\mu} \sum_{n=1}^{\infty} i^n \left( i\alpha_n \mathbf{M}_{emn}^{(3)} - \beta_n \mathbf{N}_{omn}^{(3)} + \alpha_n \mathbf{M}_{omn}^{(3)} - i\beta_n \mathbf{N}_{emn}^{(3)} \right), \tag{S5}$$

similarly, the internal fields are given by,

$$\mathbf{E}^{in} = \sum_{n=1}^{\infty} i^n \left( \delta_n \mathbf{M}_{omn}^{(1)} - i\gamma_n \mathbf{N}_{emn}^{(1)} + i\delta_n \mathbf{M}_{emn}^{(1)} - \gamma_n \mathbf{N}_{omn}^{(1)} \right), \mathbf{H}^{in} = \frac{k}{i\omega\mu} \sum_{n=1}^{\infty} i^n \left( \delta_n \mathbf{N}_{omn}^{(1)} - i\gamma_n \mathbf{M}_{emn}^{(1)} + i\delta_n \mathbf{N}_{emn}^{(1)} - \gamma_n \mathbf{M}_{omn}^{(1)} \right), \tag{S6}$$

where, $\alpha_n$, $\beta_n$, $\delta_n$ and $\gamma_n$ represent the unknown coefficients corresponding to scattered and internal fields, which are to be determined by numerically solving the boundary conditions at the interface[4], whereas the spheroidal vector wave functions $\mathbf{M}_{emn}^{(j)}$ and $\mathbf{N}_{emn}^{(j)}$ can be obtained by solving the Helmholtz equation in spheroidal coordinates using separation of variables[3]. Under far-field approximation, the asymptotic forms of the third kind of spheroidal vector wave functions are given as.

$$M_{\substack{e\\o} mn,\xi}^{(3)} = \mp(-i)^{n+1} m \cos\theta \left(\frac{d}{2}\right)^2 S_{mn}(\cos\theta) \left(\frac{1}{kr^3}\right) e^{ikr} \times_{-\cos}^{\sin} m\phi, \tag{S7}$$

$$M_{\substack{e\\o} mn,\eta}^{(3)} = (-i)^{n+1} m \frac{S_{mn}(\cos\theta)}{\sin\theta} \frac{1}{kr} e^{ikr} \times_{-\cos}^{\sin} m\phi, \tag{S8}$$

$$M_{\substack{e\\o} mn,\phi}^{(3)} = -(-i)^{n+1} \frac{dS_{mn}(\cos\theta)}{d\theta} \frac{1}{kr} e^{ikr} \times_{\sin}^{\cos} m\phi, \tag{S9}$$

similarly,

$$N_{\substack{e\\o} mn,\xi}^{(3)} = (-i)^{n+1} \left( \frac{m^2 S_{mn}(\cos\theta)}{\sin^2\theta} - \cot\theta \frac{dS_{mn}(\cos\theta)}{d\theta} - \frac{d^2 S_{mn}(\cos\theta)}{d\theta^2} \right) \left(\frac{1}{kr}\right)^2 e^{ikr} \times_{\sin}^{\cos} m\phi, \tag{S10}$$

$$N^{(3)}_{\substack{e\\o}mn,\eta} = -(-i)^n \left(\frac{dS_{mn}(\cos\theta)}{d\theta}\right)\left(\frac{1}{kr}\right) e^{ikr} \times^{\cos}_{\sin} m\phi, \tag{S11}$$

$$N^{(3)}_{\substack{e\\o}mn,\phi} = -(-i)^n m \left(\frac{S_{mn}(\cos\theta)}{\sin\theta}\right)\left(\frac{1}{kr}\right) e^{ikr} \times^{\sin}_{-\cos} m\phi. \tag{S12}$$

Where the subscript $\xi$, $\eta$ and $\phi$ corresponds to the spheroidal coordinates, whose unit vectors are correlated to the unit-vectors in spherical coordinates as, $\hat{\boldsymbol{\xi}} \to \hat{\mathbf{r}}$, $\hat{\boldsymbol{\eta}} \to -\hat{\boldsymbol{\theta}}$ and $\hat{\phi} \to \hat{\phi}$. Using the above spheroidal vector wave functions and the scattered field coefficients obtained from the boundary conditions, the scattered field and, hence, the Poynting vector $\mathbf{S}$ corresponding to the scattered field can be computed. Using equation S1, the far-field PSHS can be obtained, which is given in explicit form as follows:[3]

$$\boldsymbol{\Delta}_{\mathrm{SH}}(\theta) = \frac{1}{k} \times \mathrm{Re}\left[\frac{1}{i(\theta)}\left(\sum_{n=1}^{\infty} \alpha_n \left(\frac{\sigma_n}{\sin(\theta)} - \frac{1}{\tan(\theta)}\chi_n - \frac{d\chi_n}{d\theta}\right) \times \sum_{n=1}^{\infty}(\sigma_n\alpha_n + \chi_n\beta_n)^* + \sum_{n=1}^{\infty}\left(\beta_n\left(\frac{\sigma_n}{\sin(\theta)} - \frac{1}{\tan(\theta)}\chi_n - \frac{d\chi_n}{d\theta}\right)\right)^* \times \sum_{n=1}^{\infty}(\chi_n\alpha_n + \sigma_n\beta_n)\right)\right]. \tag{S13}$$

Here, $\sigma_n$ and $\chi_n$ are the angular functions, which are given as $S_{1n}\cos(\theta)/\sin(\theta)$ and $d(S_{1n}\cos(\theta))/d\theta$, respectively, whereas $i(\theta)$ as demonstrated by Eq. (S2) is the far-field scattering intensity given as,

$$i(\theta) = \left|\sum_{n=1}^{\infty}\left(\alpha_n\chi_n(\theta) + \beta_n\sigma_n(\theta)\right)\right|^2 + \left|\sum_{n=1}^{\infty}\left(\alpha_n\sigma_n(\theta) + \beta_n\chi_n(\theta)\right)\right|^2, \tag{S14}$$

The scattering intensity naturally arises in the definition of the PSHS irrespective of the scatterer's morphology. It is crystal clear from the above relation that the resulting PSHS is inversely related to the far-field scattered intensity. This behavior can be observed for both conventional and superscattering particles in Fig. S1. This is the fate of PSHS at the scale of a single particle, where the simultaneous enhancement of the scattered field intensity and PSHS cannot be realized even under normal incidence, as demonstrated in the earlier section. That is, the enhanced PSHS at a certain scattering angle is naturally tied to vanishingly small far-field scattering intensity and vice versa. It is important to note that the PSHS is not only a function of scattered angle but also depends upon the optical properties of the scatterer[2]. The above equation reduces to that of the spherical particle when the aspect ratio is 1, while ignoring the fast-decaying terms $d_r^{11}$ in the dipolar approximation (considering only the electric and magnetic dipoles).

**S3. Relation of spin Hall shift and scattering intensity with the scattering coefficients:**

The interference between the multipoles plays an important role in enhancing PSHS[5, 6]. Due to mode coupling, the contribution to scattering from one or several multipoles can be boosted. Therefore, the amplitude of the PSHS in the perceived location can be alter by the number of multipole terms in the expansion of scattered field, especially when the contributions from several multipoles to scattering are enhanced. From equation S13, it is obvious that the far-field intensity is in proportion with the amplitude of the coefficients of the scattered field. The simultaneous enhancement of the PSHS and the intensity requires the scattered coefficients to overlap at their resonances rather than at their tails[3]. Such a scenario is realized in conventional quasi-dual scatterer when the condition $\alpha_1 \approx \beta_1$ is satisfied, which enforces high forward directional scattering in the forward direction, i.e., when the dipolar modes constructively interfere in the forward direction. The destructive interference in the backscattered direction leads to an enhancement of the PSHS at the tail of the scattering angle.

**S4. The scattering cross-section:**

The scattering cross-section of a scatterer immersed in a non-absorbing host medium is defined as $\sigma_{\text{sca}} = \lim_{r\to\infty} 4\pi r^2 |\mathbf{E}_\text{s}|^2 / |\mathbf{E}_\text{i}|^2$, where $r$ is the radial distance, $\mathbf{E}_\text{s}$ is the scattered electric field and $\mathbf{E}_\text{i}$ is the incident electric field. By using the expansion of scattered and incident fields in the spheroidal vector wave functions, one obtains the following relation of the scattered field[4].

$$\sigma_{\text{sca}} = \frac{\lambda^2}{4\pi} \sum_{n=1}^{\infty} \left( \Pi_{nn'}^{m} \operatorname{Re}\left( |\alpha_{mn'}|^2 + |\beta_{mn'}|^2 \right) \right). \tag{S15}$$

$$\sigma_{\text{sca}} = \frac{\lambda^2}{\pi} \left| \sum_{n=1}^{\infty} \left[ \beta_n \frac{S_{1n}(\cos\theta)}{\sin\theta} + \alpha_n \frac{d[S_{1n}(\cos\theta)]}{d\theta} \right] \right|^2 \sin^2\phi + \left| \sum_{n=1}^{\infty} \left[ \beta_n \frac{d[S_{1n}(\cos\theta)]}{d\theta} + \alpha_n \frac{S_{1n}(\cos\theta)}{\sin\theta} \right] \right|^2 \cos^2\phi. \tag{S16}$$

The total cross-section of the spheroidal obtained so far is not a simple equation, due to the presence of the function $\Pi_{nn'}^{m}$, however, it is well-known that for spherical symmetric scatterer, the total contribution from a single channel cannot exceed the single channel limit, that is the maximum cross-section according to the current established limit is defined as, $\sigma_{\text{sca}}^{\max} = (2n+1)\lambda^2 / 2\pi$. With $n$ being the expansion coefficient. We normalized the scattering cross-section of each channel by the single-channel limit as given in the main text.

**S5. Demonstration of Spin Orbit interaction in the Near-Field:**

Spin and orbital angular momentum are the fundamental dynamical properties associated with optical fields that align with the propagation direction and remain uncoupled in free-space. During scattering, a stark contrast of optical constants at the interface of air and a high-index non-magnetic dielectric particle, a sharp potential gradient arises which mediates strong coupling of the polarization and spatial degree of freedom. Spin-orbit interaction in the near-field is considered the origin of PSHS. To characterize this near-field spin-orbit interaction, Maxwell's equations can be reformulated in a Dirac-like framework using optical potential $V$ and Dirac matrices $\hat{\boldsymbol{\alpha}}_i$ ($i$=1,2,3) and $\hat{\boldsymbol{\beta}}$. The concise Dirac-like equation is given below[7, 8, 9].

$$c(\hat{\boldsymbol{\alpha}} \cdot \hat{\boldsymbol{p}})\Psi + \hat{\boldsymbol{\beta}}\boldsymbol{V}\Psi = i\partial\Psi / \partial t, \qquad \text{(S17)}$$

where $\Psi = (4\omega)^{-1/2}(\varepsilon_0^{1/2}\mathbf{E}, \mu_0^{1/2}\mathbf{H})^T$ denotes the wavefunction constructed by the electric and magnetic fields and $\hat{\boldsymbol{p}}$ is the canonical momentum operator, whose expectation value $\langle \Psi | \hat{\boldsymbol{p}} | \Psi \rangle$ is directly related with the orbital momentum density $\mathbf{p}^o$[8].

$$\mathbf{p}^o = (4\omega)^{-1} \operatorname{Im}\left[\varepsilon_0 \mathbf{E} * (\nabla)\mathbf{E} + \mu_0 \mathbf{H} * (\nabla)\mathbf{H}\right]. \qquad \text{(S18)}$$

The orbital momentum density defines the trajectory of photons as optical currents, and provides a better intuition on how the photons evolve therein. On the other hand, the local spin density is related to $\langle \Psi | \hat{\boldsymbol{\alpha}} | \Psi \rangle$. Thus, the first term in equation S17 demonstrates the coupling of both spin and orbital angular momentum. Upon spin-orbit interaction in the near-field, the field losses its spin as a result of transformation to orbital angular momentum and consequently, the field near the scatterer changes its polarization, which differs from that of the incident light and far-field. This is only true in the regions of strong spin-orbit interaction. Thus, polarization analysis of the field in the vicinity of the particle would directly probe the spin-orbit coupling. The circular polarization degree in terms of total field is given as follows[7]:

$$\mathrm{CPD} = |\varepsilon_0 \mathbf{E}^* \times \mathbf{E} + \mu_0 \mathbf{H}^* \times \mathbf{H}| / (\varepsilon_0 |\mathbf{E}|^2 + \mu_0 |\mathbf{H}|^2) . \tag{S19}$$

**S6. Simulations Methods:**

All the results were confirmed by numerically solving Maxwell's equations using a finite element solver, COMSOL Multiphysics. The particles were immersed in air and the refractive index that defines the optical properties of the particles was taken from the experimental data of silicon. The long axis of the particle was taken along the direction of the propagating wave, and scattering boundary conditions were used. After computing the electric and magnetic fields, all the near-field and far-field quantities were calculated using their corresponding expressions.